\documentclass{article}
\usepackage{amssymb}
\begin{document}
\textwidth = 18truecm \textheight = 23truecm
\author{
Cyriaque ATINDOGBE, Jean-Pierre EZIN and Jo\"el TOSSA  \\
The Abdus Salam International Centre For Theoretical Physics \\
34014 Trieste, {\bf Italy} \\
\and
Institut de  Math\'ematiques et de Sciences Physiques (IMSP)\footnote{\bf {Permanent Address}} \\
 B.P. 613, Porto-Novo, {\bf B\'enin}\\
 e-mail: jtoss@syfed.bj.refer.org}
\title{Reduction of the Codimension for Lightlike Isotropic Submanifolds }
\newtheorem{theor}{Theorem}
\newtheorem{coro}{Corollary}
\newtheorem{propos}{Proposition}
\newtheorem{lem}{Lemma}
\maketitle

\begin{abstract}
We give a sufficient  condition for a lightlike isotropic
submanifold $M$, of dimension $n$, which is not totally geodesic
in a semi-Riemannian manifold of constant curvature $c$ and of
dimension $n+p\,\,\,\,(n < p)$, to admit a reduction of
codimension. We show that this condition is a necessary and
sufficient condition on the first transversal space of $M$. There
are basic and non-trivial differences from the Riemannian case,
as developed by Dajczer \textit{et al} in (\cite{dajczer}), due to the
degenerate metric on $M$. This result extends in some sense,the
one in \cite{keti} and \cite{dajczer} to lightlike isotropic submanifolds. \\

\noindent
\textit{2000 MSC, Primary : 53C50, Secondary : 53B30}\\
\noindent
\textit{Keywords :} Semi-Riemannian manifolds, Lightlike isotropic submanifolds, Screen distribution on degenerate submanifold, Reduction of codimension
\end{abstract}

\section{Introduction}
A natural generalization of the pioneering work by Gauss in
differential geometry was the study of submanifolds $f : M^n
\longrightarrow~\mathbb{R}^{n+p}$, of arbitrary codimension $p$
immersed into Euclidean $(n+p)$-spaces. An extensive work has been devoted
to these submanifolds and many results are now referred to as
classical ones on their geometrical structure. Mainly the case in
which the induced metrics on $M$ are non degenerate are examined
for instance in ( \cite{dajczer}, \cite{chen} \cite{ns},
\cite{lsz}) and references therein.

In a recent past, the growing importance of lightlike submanifolds
in global Lorentzian geometry and their use in general relativity,
motivated the study of degenerated submanifolds in a
semi-Riemannian manifold. Due to the degeneracy of the metric,
basic differences occur between the study of lightlike
submanifolds and the classical theory of Riemannian as well as
semi-Riemannian submanifolds (\cite{bejancu}, \cite{avner},
\cite{joe}).

In a point of view of physics, the idea that the Universe we live
in can be represented as a 4-dimensional hypersurface embedded in
a (4+d)-dimensional space-time manifold has attracted the
attention of many physicists. The embedding of exact solutions of
Einstein equations into higher dimensional semi-Euclidean space
can give a more adequate picture and a better understanding of
their intrinsic geometry. Higher dimensional semi-Euclidean
spaces should provide theoretical framework in which the
fundamental laws of physics may appear to be unified, as in the
Kaluza-Klein scheme, which takes into account the mutual
interaction between matter and metric (\cite{avner}, \cite{fronsdall}). \\

From the point of view of mathematics, methods and results of
submanifolds study in differential geometry might be revisited with
a greater interest to the context of degeneracy. Sometimes they
drastically change from non degenerate metric case to the degenerate metric one. As far as we know a few
literature is available on the theory of lightlike submanifolds
in semi-Riemannian manifolds. The basic work seems to be the
series by A. Bejancu and K. Duggal (\cite{bejancu}) and also D.
N. Kupeli (\cite{kupel}).\\

In this paper, generalizing earlier results in (\cite{keti}, and \cite{dajczer}), we give sufficient condition for a lightlike isotropic submanifold of dimension $n$, which is not totally
geodesic in a semi-Riemannian manifold of constant curvature $c$
and of dimension $n+p \,\,\, (n < p)$, to admit a reduction of
codimension i.e. to be immersed into an $(n+q)$-dimensional totally
geodesic submanifold of constant curvature, with $q < p$. Our main
results stand as follows
\begin{theor} Let $f : M^n \longrightarrow {\bar{M}}^{n+p}_c$ be an isometric immersion of an isotropic submanifold
$\left(M, g, S(TM^{\perp})\right)$ into a complete and simply connected
semi-Riemannian manifold with  constant sectional curvature $c$,
$\left( {\bar{M}}^{n+p}_c , \bar{g} \right)$. Suppose that :
\begin{enumerate}
\item The transversal connection $\nabla^t$ on $M^n$ is metric,
\item There exists a screen transversal subbundle $P$ of
$S(TM^\perp)$ of constant rank $q \quad (q < p)$, parallel w.r.t. the connection $\nabla^s$ on $S(TM^\perp)$, such that
$$T_1(x) \subset P(x) \quad \forall x \in M $$
\end{enumerate}
where $T_1(x)$ is the first transversal space of $f$ at $x \in M$.

Then the codimension of $f$ can be reduced to $q$
\end{theor}

The isometric immersion $f$ is said to be 1-regular if the
dimension of the transversal space is constant along $M$, and
this notion is independent of the metric of $M$. In this case, the substantial
codimension (\cite{dajczer}, p.54), or the embedding class of $M$
(\cite{avner} \cite{joe}) is the lowest value of $q$. We show that the substantial codimension of $M^n$ is equal to the
rank of its first transversal space $T_1(x)$ when the latter is of constant rank $q_0$ on $M^n$.
We have the
\begin{theor}
Let $\left(M^n , g, S(TM^\perp)\right)$ be an isometric immersion of an isotropic
non totally geodesic submanifold in $\bar{M}_{c}^{n+p}$, $(n < p)$. Then the subbundle $T_1$ is parallel w.r.t. the connection $\nabla^s$ on $S(TM^\perp)$.
\end{theor}

The paper is organized as followed. In the first paragraph, we
summarize notations and basic formulas concerning geometric
objects on lightlike submanifolds, using notations of
(\cite{bejancu}). Paragraph 2 gives the set up necessary for the
proof of the theorems and paragraph 3 gives the proofs. An
appendix shows a motivating example to illustrate the purpose of the paper.

\section{Preliminaries and basic facts}
\subsection{The General set up}
The fundamental difference between the theory of lightlike (or
degenerate) submanifolds $(M^n , g)$, and the classical theory of
submanifolds of a semi-Riemannian manifold $({\bar{M}}^{n+p} ,
\bar{g})$ comes from the fact that in the first case, the normal
vector bundle $TM^\perp$ intersects with the tangent bundle $TM$
in a non zero subbundle, denoted $Rad(TM)$, so that
\begin{equation} \label{eq1}
Rad(TM) = TM \cap TM^\perp \neq \{0\}
\end{equation}

Given an enteger $r > 0$, the submanifold $M$ is said to be
$r$-lightlike if rank$(Rad(TM))~=~r$ everywhere.

An orthogonal complementary vector subbundle of $Rad(TM)$ in $TM$
is a non degenerate subbundle of $TM$ called a screen
distribution on $M$ and denoted $S(TM)$. We have the following
splitting into an orthogonal direct sum
\begin{equation} \label{eq2}
TM = S(TM) \perp Rad(TM).
\end{equation}
From equation (\ref{eq1}), we can consider a complementary vector
subbundle $S(TM^\perp)$ of $Rad(TM)$ in $TM^\perp$. It is also a
non degenerate subbundle with respect to the metric $\bar{g}$,
and we have
\begin{equation}\label{eq3}
TM^\perp = Rad(TM) \perp S(TM^\perp).
\end{equation}

The subbundle $S(TM^\perp)$ is a screen transversal vector bundle
of $M$. The subbundle $S(TM)$ being non degenerate, so is
$\left(S(TM)\right)^\perp$ and the following holds
\begin{equation} \label{eq4}
T{\bar{M}}|_M = S(TM) \perp \left(S(TM)\right)^\perp.
\end{equation}
Note that $S(TM^\perp)$ is a subbundle of $\left(S(TM)\right)^\perp$ and, since
both are non degenerate, we have
\begin{equation} \label{eq5}
\left(S(TM)\right)^\perp = S(TM^\perp) \perp
\left(S(TM^\perp)\right)^\perp
\end{equation}
One frequently denotes a lightlike submanifold $M$ by $\left(M, S(TM), S(TM^\perp) \right)$
to refer to the above subbundles. \\

In fact, $Rad(TM)$ is a subbundle of $\left(S(TM^\perp)\right)^\perp$. Let $ltr(TM)$
denote its complementary vector bundle in $\left(S(TM^\perp)\right)^\perp$.
One has
$$\left(S(TM^\perp)\right)^\perp = Rad(TM) \oplus ltr(TM) $$
The subbundle $ltr(TM)$ is called a lightlike transversal vector
bundle of $M$. The subbundle $tr(TM)$ defined by
$$tr(TM) = ltr(TM) \perp S(TM^\perp)$$
is called a transversal vector bundle of $M$ and plays an
important role in the study of the geometry of lightlike
submanifolds. We always have $tr(TM) \cap TM^\perp
\neq tr(TM)$. That is $tr(TM)$ is never orthogonal to $TM$. \\

From now on, given a vector bundle $E$, we denote $\Gamma(E)$ the
space of smooth sections of $E$. 

Summarizing the above statements, we have the following decomposition
\begin{eqnarray} \label{eq6}
T{\bar{M}}|_M & = & TM \oplus tr(TM) \nonumber \\
              & = & S(TM) \perp S(TM^\perp)\perp \left(Rad(TM) \oplus ltr(TM) \right),
\end{eqnarray}
which gives rise to a local quasi-orthonormal field of frames on $\bar{M}$ along
$M$ (see \cite{bejancu}) denoted by $\left( \xi_i, N_i , X_a , W_\alpha \right)$,
where
\begin{enumerate}
\item $\{\xi_i\}$ and $\{N_i\}, \,\,\, i \in \{1, \cdots , r\}$ are lightlike
basis of $\Gamma\left(Rad(TM)|_{\cal U}\right)$ and
$\Gamma\left(ltr(TM)|_{\cal U}\right)$ respectively;
\item $\{X_a\}, \,\,\, a \in \{r+1, \cdots , m \}$ is an orthonormal basis of
$\Gamma\left(S(TM)|_{\cal U}\right);$
\item $\{W_\alpha\}, \,\,\, \alpha \in \{r+1, \cdots , n\}$ an orthonormal basis
of $\Gamma\left(S(TM^\perp)|_{\cal U}\right),$
\end{enumerate}
relative to a coordinate neighborhood ${\cal U} \subset M$. \\

A lightlike submanifold is said to be \textit{isotropic} if $Rad(TM) =
TM$. In this case, we deduce from (\ref{eq2}) that $S(TM) =
\{0\}$. This requires that $n < p$ and the formula (\ref{eq6})
reduces to
\begin{equation} \label{eq7}
T{\bar{M}}|_M  =  TM \oplus tr(TM)
               =  S(TM^\perp) \perp \left(Rad(TM) \oplus ltr(TM) \right)
\end{equation}

In the sequel, the lightlike submanifold $M$ is supposed to be isotropic.

\subsection{Induced connections}
Let $\bar{\nabla}$ denoted the Levi-Civita connection on $\bar{M}$ and $\nabla$ the induced
connection on $M$. For all $X, Y \in \Gamma(TM)$, and $V \in \Gamma(tr(TM))$, we deduce from (\ref{eq7}) that
\begin{equation} \label{eq8}
\bar{\nabla}_XY = \nabla_XY + h^l(X,Y) + h^s(X,Y)
\end{equation}
and
\begin{equation}\label{eq9}
\bar{\nabla}_XV = - A_VX + D^{l}_{X}V + D^{s}_{X}V
\end{equation}
where $h^l$ and $h^s$ are $\Gamma\left(ltr(TM)\right)$-valued,
and $\Gamma\left(S(TM^\perp)\right)$-valued respectively. They
are called  the lightlike and the screen second fundamental forms
of $M$, respectively. As usual, $A_V$ denotes the shape operator with respect to $V$. \\

The second fundamental form of $M$ with respect to $tr(TM)$ is defined by
\begin{equation} \label{eq10}
h(X, Y) = h^l(X, Y) + h^s(X, Y), \quad X, Y \in \Gamma(TM)
\end{equation}

Let $L$ and $S$ denote the projection morphism of $tr(TM)$ on
$ltr(TM)$ and $S(TM^\perp)$ respectively. In (\ref{eq9}) we have
$$D^{l}_{X}V = L \left(\nabla^{t}_{X}V \right), \quad D^{s}_{X}V =
S \left(\nabla_{X}^{t}V \right), \quad \forall X \in \Gamma(TM),
\quad \forall V \in \Gamma\left(tr(TM)\right) $$ where $\nabla_{X}^{t}$ stands for
the transversal linear connection on $M$. The transformations
$D^l$ and $D^s$ do not define linear connections on $tr(TM)$
(\cite{bejancu}, p.27), but define two Otsuki connections on
$tr(TM)$ with respect to the vector bundle
morphisms $L$ and $S$. \\

Since the submanifold $M$ is isotropic, the lightlike second
fundamental form $h^l$
vanishes identically on $M$ (\cite{bejancu}, p.157). \\

Define the $C^\infty(M)$-bilinear mappings, $D^l$ and $D^s$ by
$$\matrix{
D^l : \Gamma(TM) \times \Gamma\left(S(TM^\perp)\right) &
\longrightarrow & \Gamma\left(ltr(TM)\right) \cr
                                        (X, SV)        & \mapsto         & D^l(X, SV) = D^{l}_{X}(SV)
                                        }$$
and
$$\matrix{
D^s : \Gamma(TM) \times \Gamma\left(ltr(TM)\right) & \longrightarrow & \Gamma\left(S(TM^\perp)\right) \cr
                                        (X, LV)    & \mapsto         & D^s(X, LV) = D^{s}_{X}(LV)
                                        }$$
Then we have
\begin{equation} \label{eq11}
\bar{\nabla}_{X}N = - A_{N}X + \nabla_{X}^{l}N + D^s(X,N)
\end{equation}
\begin{equation} \label{eq12}
\bar{\nabla}_{X}W = - A_{W}X + \nabla_{X}^{s}W + D^l(X,W)
\end{equation}
where $\nabla^s$ and $\nabla^l$ are linear connections on
$S(TM^\perp)$ and $ltr(TM)$  respectively; $X \in \Gamma(TM)$, $N
\in \Gamma\left(ltr(TM)\right)$ and
$W \in \Gamma\left(S(TM^\perp)\right)$.\\

As shown in (\cite{bejancu}, p.166) $M$ is totally geodesic if
and only if $D^l(. , W) = 0, \quad \mbox{for all} \quad  W \in
\Gamma\left(S(TM^\perp)\right)$.

 A direct computation shows that, for all $X \in \Gamma(TM), \quad V, V' \in
\Gamma\left(tr(TM)\right)$ we have
\begin{equation} \label{eq13}
\left(\nabla_{X}^{t} \bar{g}\right)(V, V') = - \left( \bar{g}(A_{V}X, V') +
\bar{g}(A_{V'}X, V) \right)
\end{equation}
so that the transversal linear connection $\nabla^t$ on $tr(TM)$ is not metric in general. \\

The first transversal space at $x \in M$ of the isometric immersion $f$ is
defined  as the subspace
$$T_1(x) = span \{h^s(X,Y), X, Y \in \Gamma(T_xM)\}$$

For the proof of theorems, we need the following two lemmas.
\begin{lem} \label{Lemma 1}
If the transversal linear connection $\nabla^t$ on $tr(TM)$ is
metric, then $\displaystyle A_W = 0$ for all $W \in \Gamma \left(
S(TM^\perp) \right)$
\end{lem}

{\bf Proof :}  Due to equation (\ref{eq13}), $\nabla^t$ is metric,
if and only if  $A_W$ is $\Gamma\left(S(TM)\right)$-valued for all $W \in \Gamma
\left( S(TM^\perp) \right)$. The lemma follows from the fact
that  $M$ being isotropic, $S(TM) = \{0\} \quad \square$.

\begin{lem} \label{Lemma 2}
 For any $x \in M$, the first transversal space $T_1(x)$ has the
characterization
\begin{equation} \label{eq14}
T_1(x) = \left\{ V = W+N \in \Gamma \left( tr(TM) \right)\,\,\,
/\,\,\, D^l(. , W) = 0 \right\}^\perp
\end{equation}
\end{lem}

{\bf Proof :} Because $M$ is a non totally geodesic isotropic
submanifold of $\bar{M}$, Lemma 2 shows that $T_1$ is not
trivial, that is $T_1(x) \neq \{0\}$, for all $x \in M$. \\
 Let $V= h^s(X,Y), \quad X, Y \in \Gamma \left( (TM) \right)$, be a
generic  element of $T_1(x)$ and $U \in A(x)^\perp$ with
$$A(x) := \left\{ V = W+N \in \Gamma \left( tr(TM) \right) \quad / D^l(. , W) = 0 \right\}^\perp$$
then

\begin{eqnarray*}
\bar{g}(U, V) & = & \bar{g}\left(h^s(X, Y), W+N \right) \\
              & = & g \left( A_{W}X , Y \right) - \bar{g} \left(Y, D^l(X, W) \right) \\
              & = & 0
\end{eqnarray*}
where we use Lemma 1 and the definition of $A(x)$. Thus,
\begin{eqnarray*}
V \in T_1(x) & \Longleftrightarrow & \bar{g}(V, U) = 0 \quad \forall \,U \in A(x)^\perp \\
             & \Longleftrightarrow & V \in {\left( A(x)^\perp \right) }^\perp =
             A(x),
\end{eqnarray*}
so $T_1(x) = A(x)\quad \square$\\

\section{Proof of Theorems}
\subsection{Proof of Theorem 1}
First of all, note that $P$ is a $\nabla^s$-parallel subbundle of constant
rank $q$ of the bundle $S(TM^\perp)$ implies that
$$\nabla^{s}_{X}W \in P, \quad \forall X \in \Gamma (TM), \quad \forall W \in \Gamma(P) $$

Then consider as usual the three cases $c=0, \quad c
> 0, \quad c < 0$.\\

\noindent
{\bf Case  $\mathbf{c = 0}$}

For $x_0 \in M$, we prove that $f(M) \subset T_{x_0}M \oplus
P(x_0)$. Let $\mu$ be a section of the complementary orthogonal
bundle of $P$ in $S(TM^\perp)$, $\gamma : I \longrightarrow M$ a
regular curve on $M$ and $\mu_t$ the parallel transport
of $\mu$ along $\gamma$. \\

Since $P$ is parallel in $\Gamma \left( S(TM^\perp) \right)$, so
is its orthogonal complementary $P^\perp$ in the subbundle $\Gamma
\left( S(TM^\perp) \right)$ and
$$\mu_t = \nabla_{\gamma'}^{s}\mu \in  \Gamma\left(P^{\perp}_{\gamma (t)}\right) , \quad \forall t \in I $$
Using Weingarten formula, we have
$${\bar{\nabla}}_{\gamma '}\mu_t = - A_{\mu_t}\gamma' + D^l(\gamma' , \mu_t ) + \nabla_{\gamma'}^{s}\mu_t$$
But
$$\mu_t \in \Gamma\left(P^{\perp}_{\gamma(t)}\right) \subset \Gamma\left(S(TM^\perp)\right), \quad \forall t \in I$$
Lemma 1 yields $A_{\mu_t}^{\gamma'} = 0$ for all $t \in I$. \\
Moreover, $ \displaystyle \mu_t \in P^{\perp}_{\gamma(t)} \subset
T_{1}(\gamma(t)) $ from assumption of the theorem. So Lemma 2
gives
$$D^l(\gamma', \mu_t) = 0 \quad \forall\,t \in I $$

And because $\mu_t$ is the parallel transport in $P^\perp$ of
$\mu$ along $\gamma$,
we have $\nabla_{\gamma'}^{s}\mu_t = 0$ for all $t$ in $I$. \\

We deduce that $\bar{\nabla}_{\gamma'}\mu_t = 0$ for all $t \in I$, so that
$\mu_t = \mu$ is a constant vector in $\mathbb{R}^{n+p}$. \\
Hence
\begin{eqnarray*}
\frac{d}{dt}\bar{g} \left( f(\gamma(t)) - f(x_0), \mu_t \right) & = & \bar{g} \left( f_{\ast}\gamma'(t), \mu \right) \\
                                                                & = & 0
\end{eqnarray*}
We conclude that
$$\bar{g}\left(f(\gamma(t)-f(x_0), \mu \right) = 0 \quad \forall t \in I$$
and
$$f(\gamma(t)) - f(x_0) \in \left( P^{\perp}_{\gamma(t)} \right)^\perp = P_{\gamma(t)},
\quad \forall t \in I$$

Due to the fact that $\gamma$ and $\mu$ are arbitrary on $M$, we have
$$f(M) \subset T_{x_0}(M) \oplus P(x_0) \cong \mathbb{R}^{n+q}$$
which is a totally geodesic $(n+q)$-dimensional subspace of $\mathbb{R}^{n+p}$ \\

{\bf Case $\mathbf{c > 0}$}

The isotropic submanifold $M^n$ is isometrically immersed into a
pseudosphere $S_{c}^{n+p}$. Consider the isometric immersion
$$\tilde{f} = i \circ f : M^n \longrightarrow \mathbb{R}^{n+p+1}$$
where the map $i$ is the natural injection of $S^{n+p}_{c}$ into
$\mathbb{R}^{n+p+1}$. Then $$tr(\tilde{T}_x M) = tr(T_xM) \oplus
<f(x)>$$ with
$$<f(x)>\,\, \subset \,\, S\left((\tilde{T}_xM^\perp)\right)$$
where $< f(x) > := Span\{f(x)\}$. \\

We deduce that
$${\tilde{T}}_1(x) \subset T_1(x) \oplus <f(x)> \subset P(x) \oplus <f(x)> = \tilde{P}(x)$$
And then
$$\tilde{T}_1(x) \subset S(T_{x}M^\perp)  \oplus  <f(x)> = S(\tilde{T}M^\perp), \quad \forall x \in M$$

The orthogonal complementary of $\tilde{P}(x)$ in
$S(\tilde{T}M^\perp)$ and of $P(x)$ in $S(TM^\perp)$, which is
parallel w.r.t. the transversal screen connection $\nabla^s =
\tilde{\nabla}^{s}|_{S(TM^\perp)}$, are equal, and
$${\tilde{\nabla}}^{s}_{X}W = \tilde{D}^{s}_{X}W = D^s(X,W), \quad \forall W \in
\Gamma \left(\tilde{T}(S(TM^\perp))\right)$$ Thus
\begin{eqnarray*}
\bar{g}\left( \tilde{\nabla}^{s}_{X}{\tilde{f}(x)}, W \right) & = & \bar{g}\left( \bar{D}^{s}_{X}\tilde{f}(x), W \right) \\
& = & X.{\bar{g}}\left( \tilde{f}(x), W \right) - \bar{g}\left(\tilde{f}(x), \nabla_{X}^{s}W \right) \\
& = & 0
\end{eqnarray*}
and therefore
$$\tilde{\nabla}^{s}_{X}{\tilde{f}(x)}  \in  <f(x)>$$

and $< f(x) >$ is a transversal  vector subbundle who is parallel
w.r.t. the connection $\tilde{\nabla}^s$. We conclude that
$\tilde{P}$ is parallel w.r.t. $\tilde{\nabla}^s$, and as in the
case $c=0$, we have
$$\tilde{f}(M) \subset \tilde{T}_{x_0}M \oplus \tilde{P}(x_0) = T_{x_0}(M) \oplus
P(x_0) \oplus <f(x_0)> \cong \mathbb{R}^{n+q+1}$$

So $f(M) \subset S^{n+p}_c \cap \mathbb{R}^{n+q+1} = S_{c}^{n+q}$ which is totally
geodesic in $S_{c}^{n+p}$. This proves the case $c > 0$. \\

{\bf Case $\mathbf{c < 0}$}

 The general scheme holds as for $c > 0$. Now $\tilde{f}$ maps
$M^n$ into $\mathbb{L}^{n+p+1}$, the Lorentzian space $\mathbb{R}_{1}^{n+p+1}$ and
we get that
$$\tilde{f}(M) \subset \tilde{T}_{x_0}M \oplus \tilde{P}(x_0) = T_{x_0}(M) \oplus
P(x_0) \oplus <f(x_0)>$$ where $f(x)$ is spacelike. Then
$$\tilde{f}(M) \subset  \mathbb{L}^{n+q+1}$$
and
$$f(M) \subset \mathbb{H}^{n+p}_c \cap \mathbb{L}^{n+q+1} \cong \mathbb{H}_{c}^{n+q}$$
and $M$ admits a reduction of codimension, which completes the proof $\square$. 

\subsection{ Proof of Theorem 2}  We have $T_1(x) \subset \left( S(TM^\perp)
\right), \quad \forall x \in M$. To prove that $T_1$ is parallel,
we will prove that its orthogonal complementary in $\Gamma \left( S(TM^\perp) \right)$ is parallel. \\
So, if $\eta \in T^{\perp}_{1}$, we have to prove that
$$\nabla_{Z}^{s}\,\eta  \in  T_{1}^{\perp}, \quad \forall Z \in \Gamma
\left(TM) \right)$$
i.e.
\begin{equation} \label{eq15}
\bar{g}\left( h^s(X,Y), \nabla_{Z}^{s}\eta \right) = 0 \quad \forall X, Y, Z \in \Gamma(TM)
\end{equation}
Set
$$ \eta = N + W , \quad N \in \Gamma\left (ltr(TM) \right)\,\, , \,\, \quad  W \in
\Gamma\left( S(TM^\perp) \right); $$
 then
$$\bar{g}\left( h^s(X,Y), \nabla_{Z}^{s}\eta \right) = g \left(
A_{\nabla_{Z}^{s}} \eta , Y \right) - \bar{g}\left(Y, D^l(X, W)\right)$$
But using Lemma 1, we have
$$\nabla_{Z}^{s}\eta \in \Gamma \left( S(TM^\perp) \right) \Longrightarrow A_{\nabla_{Z}^{s}\eta} = 0$$
and
$$\eta = N+W \in T_{1}^{\perp} \Longrightarrow D^l(X, W) = 0$$
We deduce that
$$\bar{g} \left( h^s(X,Y), \nabla_{Z}^{s}\eta \right) = 0 \quad \forall X, Y, Z \in \Gamma(TM)$$
and then $\nabla_{Z}^{s}\eta \in T_{1}^{\perp}$ so that $N_{1}^{\perp}$ is
parallel w.r.t.
the connection $\nabla^s$. This proves the theorem 2 $\square$. \\

As a consequence of the two theorems, we have the following

\begin{propos}
A necessary and sufficient condition for the isotropic immersion $f : M^n \rightarrow {\tilde{M}}^{n+p}_{c}, \quad n < p$ to admit a reduction of codimension, is that the isotropic immersion is 1-regular of constant rank $q$, and the substantial codimension is $q$.
\end{propos}

\appendix
\section*{Appendix}
These ideas are illustrated through the following example. \\

Suppose $M$ is a surface of $\mathbb{R}^{5}_{2}$, Euclidean space
$\mathbb{R}^5$ with a semi-Euclidean metric $\bar{g} = diag(-1,
-1, +1, +1, +1)$, given by equations
$$x^1 = \frac{1}{\sqrt 2}\left(x^4 + \sinh x^5 \right) ; \quad
x^2 = \frac{1}{\sqrt 2}\left(x^4 - \sinh x^5 \right) ; \quad x^3
= \cosh x^5 $$ and set $\left( u = x^4 , v = x^5 \right)$ a system
of coordinate on $M$. We derive the following
$$TM = Span \left\{ \xi_1 , \xi_2 \right\}$$
with
$$\xi_1 = \frac{\partial}{\partial u} = \frac{1}{\sqrt 2}\frac{\partial}{\partial x^1}
+ \frac{1}{\sqrt 2}\frac{\partial}{\partial x^2} + \frac{\partial}{\partial x^4} ; \quad
$$
$$\xi_2 = \frac{\partial}{\partial v} = \frac{\cosh x^5}{\sqrt 2}\frac{\partial}{\partial x^1} -
\frac{\cosh x^5}{\sqrt 2}\frac{\partial}{\partial x^2} + \sinh x^5 \frac{\partial}{\partial x^3} +
\frac{\partial}{\partial x^5}$$
and
$$TM^{\perp} = Span \left\{U_1 = \xi_1, \quad U_2=\xi_2, U_3 =\frac{\partial}{\partial x^3}+
\frac{1}{\sqrt 2}\frac{\partial}{\partial x^4} - \sinh x^5 \frac{\partial}{\partial x^5} \right\}$$
\\
It follows that $Rad(TM) = TM \subset TM^{\perp}$ and $M$ is an isotropic surface
of $\mathbb{R}_{2}^{5}$. \\

The subbundle $S(TM^{\perp})$ is a complementary vector bundle of
$Rad(TM)$ in $TM^{\perp}$. We take (there is no unicity),
$$S(TM^{\perp}) = Span \left\{ W_1 = \frac{\sinh x^5}{\sqrt 2}\frac{\partial}{\partial x^1} -
\frac{\sinh x^5}{\sqrt 2}\frac{\partial}{\partial x^2} + \cosh x^5 \frac{\partial}{\partial x^3} \right\}$$
\\

{\bf Construction of $ ltr(TM)$ :}  A basis $\left\{N_1, N_2
\right)$ of $ltr(TM)$ on a coordinate neighborhood $\cal U$
satisfies :
\begin{eqnarray} \label{a1}
\bar{g}\left(N_i, N_j \right) & = & 0, \quad \forall i, j \in \{1,
2\} \nonumber \\
 \bar{g}\left(\xi_1, N_2\right) & = & \bar{g}\left(\xi_2, N_1\right) = 0  \\
\bar{g}\left(N_1, \xi_1\right) & = & \bar{g}\left(N_2,
\xi_2\right) = 1 \nonumber
\end{eqnarray}

Using (\ref{a1}) we obtain that $$ltr(TM) = Span\{N_1, N_2\},$$
with,
\begin{eqnarray*}
N_1 & = & \frac{1}{2}\left( - \frac{1}{\sqrt 2}\frac{\partial}{\partial x^1} -
\frac{1}{\sqrt 2}\frac{\partial}{\partial x^2} + \frac{\partial}{\partial x^4} \right) \\
N_2 & = & \frac{1}{2} \left( - \frac{\cosh x^5}{\sqrt 2}\frac{\partial}{\partial x^1} +
\frac{\cosh x^5}{\sqrt 2}\frac{\partial}{\partial x^2} - \sinh x^5 \frac{\partial}{\partial x^3} +
\frac{\partial}{\partial x^5} \right)
\end{eqnarray*}
and deduce that
$$tr(TM) = ltr(TM) \oplus Rad(TM) = Span\{W_1, N_1, N_2\}$$

A straightforward calculation gives
$$\bar{\nabla}_{\xi_1}\xi_1 = \bar{\nabla}_{\xi_1}\xi_2 = \bar{\nabla}_{\xi_2}\xi_1 = 0 \quad
\bar{\nabla}_{\xi_2}\xi_2 = W_1$$ We deduce that $M$ is not
totally geodesic in $\mathbb{R}^{5}_{2}$. \\

Moreover we have for all $X, Y, \in \Gamma(TM), \quad X =
X^{i}\xi_i, \quad   Y = Y^{j}\xi_j$,

\begin{eqnarray*}
\bar{\nabla}_{X}Y & = & {\nabla}_{X}Y + h^{s}(X, Y) \\
                  & = & \left[\left(X^{1}(\xi_{1}(Y^{1})) +
X^{2}(\xi_2(Y^1))\right))\xi_{1} + \left(X^{1}(\xi_{1}(Y^2)
+X^{2}(\xi_2(Y^2))\right)\xi_2 \right] + \\
 &  & + X^{2}Y^{2}W_1
\end{eqnarray*}

 and $$h^{s}(X, Y) = \bar{g}(X, N_2) \bar{g}(Y, N_2) W_1$$

So that
\begin{equation} \label{a2}
 h^{s}_{1}\left(\xi_2, \xi_2 \right) = 1
\end{equation}

From (\ref{a2}) we infer that
\begin{eqnarray*}
T_1(x) & = & Span\left\{h^{s}(X,Y), X, Y \in \Gamma(T_xM)\right\} \\
       & = & S\left( T_xM^\perp \right)
\end{eqnarray*}
which is of constant rank $q = 1$ for all $x \in M$. From above
and proposition , $M$ admits a reduction of codimension to 1,
that is there exists a totally geodesic 3-dimensional submanifold
of $\mathbb{R}^{5}_{2}$ into which $M$ can be isometrically
immersed.

\section*{Acknowledgement}The authors would like to thank
Professor M. Virasoro, the Abdus Salam International Centre for
Theoretical Physics (ICTP) for hospitality and the Swedish International 
Development Cooperation Agency (SIDA) for financial support
through the {\bf Associate Scheme programme} and the {\bf Young
Student programme} of the Abdus Salam ICTP.

\end{document}